\documentstyle[12pt]{article}

\def\be{\begin{equation}}
\def\ee{\end{equation}}
\def\ben{\begin{displaymath}}
\def\een{\end{displaymath}}
\def\ba{\begin{array}{c}}
\def\bal{\begin{array}{l}}
\def\ea{\end{array}}

\begin{document}

\titlepage
\vspace*{2cm}

 \begin{center}{\Large \bf
 Strengthened
  ${\bf PT}-$symmetry with ${\bf P}\neq {\bf P}^\dagger$
  }\end{center}

\vspace{10mm}

 \begin{center}
Miloslav Znojil

 \vspace{3mm}

\'{U}stav jadern\'e fyziky AV \v{C}R, 250 68 \v{R}e\v{z}, Czech
Republic\footnote{e-mail: znojil@ujf.cas.cz}

\end{center}

\vspace{5mm}



\section*{Abstract}

In Quantum Mechanics working with non-Hermitian $PT-$symmetric
Hamiltonians (i.e., with an indefinite metric $P$ in Hilbert
space) we propose to relax the usual constraint $P=P^\dagger$. We
show that this merely induces certain ``hidden" symmetries
responsible, say, for the degeneracy of levels.  Using a triplet
of the coupled square wells for illustration we show that the
bound states may remain stable in a large domain of couplings.

 \vspace{5mm}

PACS

03.65.Ca
03.65.Ge;

\newpage

\section{${\cal PT}-$symmetric Quantum Mechanics and its two alternative
scenarios}

${\cal PT}-$symmetric Quantum Mechanics of C. Bender et al
\cite{BBjmp} studies non-Hermitian Hamiltonians $H\neq H^\dagger$
with the peculiar property
 \be
 {\cal PT}\,H = H\,{\cal PT}\,.
 \label{jedna}
 \ee
In the original and simplest one-dimensional version of the theory
\cite{BB} the symbol ${\cal T}$ denotes the complex conjugation,
i.e., an antilinear involution with the property ${\cal T}\,{\rm
i}\,\partial_t\,{\cal T}= -{\rm i}\,\partial_t$ interpreted as
time reversal. The Hamiltonians themselves are assumed symmetric
so that we may replace $H \to {\bf \it h}={\bf \it h}^T$ and
deduce that ${\cal T}\,{\bf \it h}\,{\cal T} = {\bf \it h}^*\equiv
{\bf \it h}^\dagger$, ${\cal T}\,{\bf \it h}\,{\cal P}\,{\cal T} =
{\bf \it h}^\dagger\,{\cal P}^*$ and
 \be
 h^\dagger = {\cal P}^{-1}\,h\,{\cal P} = {\cal P}^*\,h\,
 \left ({\cal P}^*
 \right )^{-1}.
 \label{jednatel}
 \ee
The choice of the parity ${\cal P}$ with the properties ${\cal
P}={\cal P}^*={\cal P}^{-1}={\cal P}^T={\cal P}^\dagger$ in
ref.~\cite{BB} inspired A. Mostafazadeh \cite{ali} who recommended
a transition from the physics-inspired symmetry (\ref{jedna}) to
its mathematically equivalent representation in the form of the
${\cal P}-$pseudo-Hermiticity requirement
 \be
 H^\dagger = {\cal P}\,H\,{\cal P}^{-1}\,.
 \label{aliholk}
 \ee
He emphasized that eq.~(\ref{aliholk}) may be read as an
isospectrality property where one might work with asymmetric
Hamiltonians {\em and} with the ``generalized parity" operators
which need not be involutive at all, ${\cal P} \to {\bf P} \neq
{\bf P}^{-1}$.

In such a perspective the boldface symbol ${\bf P}$ may represent
an arbitrary auxiliary operator. The involutive character of the
Hermitian conjugation in eq.~(\ref{aliholk}) implies that
 \be
 H=\left (H^\dagger\right )^\dagger
  = \left ({\bf P}^\dagger\right )^{-1}\,H^\dagger\,{\bf P}^\dagger=
   \left ({\bf P}^\dagger\right )^{-1}\, {\bf P}\,H\,{\bf P}^{-1}\,{\bf
   P}^\dagger
   \label{symm}
   \ee
which gives us the two alternative possibilities;
\begin{description}
\item{\bf [a]}
 a self-adjoint pseudo-metric ${\bf P}={\bf
P}^\dagger$ is chosen, or
\item{\bf [b]}
non-Hermitian operators ${\bf P}\neq {\bf
P}^\dagger$ are admitted.
\end{description}
 \noindent
In the light of the current literature the ``trivial symmetry"
scenario {\bf [a]} seems to be ``the only useful" option where
${\bf P}$ becomes a pseudo-metric (often called ``indefinite
metric") in the physical Hilbert space of states ${\cal H}$.

We believe that the non-Hermitian alternative {\bf [b]} may prove
equally interesting. Indeed, there is no real reason for ignoring
the class of operators $S={\bf P}^{-1}\,{\bf P}^\dagger \neq I$
which represent a nontrivial symmetry (\ref{symm}) of the
Hamiltonian. In what follows we intend to support such a point of
view by an illustrative construction. For this purpose we shall
pick up one of the most elementary coupled-channel symmetries $S
\neq I$ derived from the non-involutive and non-Hermitian toy
operator
 \be
 {\bf P} =\left (
 \begin{array}{ccc}
 0&0&{\cal P}\\
 {\cal P}&0&0\\
 0&{\cal P}&0
 \ea
 \right )=\left ({\bf P}^\dagger \right )^{-1}=
 {\bf P}^{-2}.
 \label{provazanekanaly3}
 \ee
Provided that its sub-operator ${\cal P}$ remains defined as the
parity, ${\cal P}\psi(x) = \psi(-x)$, it cannot be interpreted as
a metric because its eigenvalues are complex.

A supplementary reason for the present use of the non-Hermitian
${\bf P}$ of eq.~(\ref{provazanekanaly3}) emerges once we return
back to the symmetric Hamiltonians $h=h^T$ in eq.~(\ref{jednatel})
where any ``early generalization" ${\cal P} \to {\cal R}$ of the
parity would lead immediately to an alternative explicit
constraint
 \be
 {\cal R}\, {\cal R}^*\,{\cal \it h}
 = {\bf \it h}\,{\cal R}\, {\cal R}^*\,.
 \ee
We see that the ``alternative generalized parities" ${\cal R}$
would have to be unitary whenever one assumes that they are not
asymmetric. One should keep in mind that the latter consistency
constraint definitely differs from its predecessor
eq.~(\ref{symm}). Of course, it can again be interpreted as
imposing an additional symmetry upon the Hamiltonian. Thus, one
feels that eq.~(\ref{aliholk}) with self-adjoint ${\cal P}$ need
not necessarily offer the only productive way towards a
generalization.

\section{Non-Hermitian triplet of coupled square wells}

Equation (\ref{provazanekanaly3}) and the non-Hermitian
Hamiltonian of the triple-channel form
 \be
 H=
 \left (
 \begin{array}{ccc}
  -\frac{d^2}{dx^2}+D_a(x)&U_b(x)&V_b(x)\\
 U_a(x)& -\frac{d^2}{dx^2}+D_b(x)&W_b(x)\\
 V_a(x)&W_a(x)& -\frac{d^2}{dx^2}+D_c(x)
 \ea
 \right )\,
 \label{vazanekanaly3}
 \ee
will be assumed inter-related by our present modification
 \be
 H^\dagger = {\bf P}\,H\,{\bf P}^{-1}\,,
 \ \ \ \ \ \ \ \ \ {\bf P}\neq {\bf P}^\dagger
 \label{aliholker}
 \ee
of the ${\bf P}-$pseudo-Hermiticity condition (\ref{aliholk})
re-written in the form adapted to the less common scenario {\bf
[b]}. Our choice of the interaction potentials will be dictated by
the exact solvability requirement in a way inspired by the
simplicity of the various {\em single-channel} square-well models
\cite{Bijan} -- \cite{Bijanb} in scenario {\bf [a]}.

No innovations will occur in the real, Hermitian part of the
present potentials,
 \be
 \bal
 {\rm Re}\,D_{a,b,c}(x) ={\rm Re}\,U_{a,b}(x) ={\rm Re}\,V_{a,b}(x)
 ={\rm Re}\,W_{a,b}(x)=0,
 \ \ \ \ \ \ \  x \in (-1,1),\\
 {\rm Re}\,D_{a,b,c}(x) ={\rm Re}\,U_{a,b}(x) ={\rm Re}\,V_{a,b}(x)
 ={\rm Re}\,W_{a,b}(x) =\infty,
 \ \ \ \ \ \ \  x \notin (-1,1).
 \ea
 \label{tri}
 \ee
In the same spirit, the imaginary potentials in Hamiltonian
(\ref{vazanekanaly3}) will be postulated piecewise constant. Their
specification
 \be
 \bal
 {\rm Im}\,U_{a,b}(x) = {\rm Im}\,V_{a,b}(x)
 = {\rm Im}\,W_{a,b}(x) =Y >0, \ \ \ \ \, \ \ \ \ \
  x \in (-1,0),\\
 {\rm Im}\,U_{a,b}(x) = {\rm Im}\,V_{a,b}(x)
 = {\rm Im}\,W_{a,b}(x) =-Y, \ \ \ \ \  \ \ \ \ \ x \in
 (0,1),\\
 {\rm Im}\,D_{a,b,c}(x) = Z, \ \ \ \ \
  x \in (-1,0),\ \ \ \ \ \ \ \ \
 {\rm Im}\,D_{a,b,c}(x) = -Z, \ \ \ \ \
  x \in (0,1)
 \ea
 \label{SQW3}
 \ee
containing two free real coupling constants results from the
pseudo-Hermiticity condition (\ref{aliholker}) and from our choice
of the generalized parity~(\ref{provazanekanaly3}). This defines
the family of the coupled-channel Schr\"{o}dinger equations
 \be
 H\,
 \left (
 \ba
 \varphi_a(x)\\
 \varphi_b(x)\\
 \varphi_c(x)
 \ea
 \right )
 =
 E\,\left (
 \ba
 \varphi_a(x)\\
 \varphi_b(x)\\
 \varphi_c(x)
 \ea
 \right )
 \label{SE}
 \ee
where the energies $E$ may be assumed real, for the small
couplings $Y$ and $Z$ at  least \cite{Langer}. As long as we put
our system in a box (\ref{tri}), the elementary boundary condition
 \be
 \left .
 \left (
 \ba
 \varphi_a(x)\\
 \varphi_b(x)\\
 \varphi_c(x)
 \ea
 \right )
 \right |_{x=\pm 1}
 =\left (
 \ba
 0\\0\\0
 \ea
 \right )
 \label{bc}
 \ee
determines all the bound states of the model.

\section{Solutions and the  degeneracy of their spectrum}

The obvious ansatz for the general solution
 \be
 \ba
 \varphi_{a,b,c}(x)= \left \{
 \begin{array}{ll}
  C_L^{(a,b,c)}\,\sin \kappa_L(x+1)
 , \ \ & x \in (-1,0), \\
  C_R^{(a,b,c)}\,\sin \kappa_R(-x+1),
  \ \ & x \in (0,1)
 \ea
 \right .
 \ea
 \label{ansatzf}
 \ee
takes into account the boundary conditions and necessitates only
the appropriate matching of all the three wave functions
$\varphi_{a,b,c}(x)$ {\em and} of their first derivatives at
$x=0$. This imposes the six complex constraints
 \ben
  C_L^{(a,b,c)}\,\sin \kappa_L=
  C_R^{(a,b,c)}\,\sin \kappa_R,
 \ \ \ \ \ \ \ \ \
   \kappa_L\,C_L^{(a,b,c)}\,\cos \kappa_L=
 -  \kappa_R\,C_R^{(a,b,c)}\,\cos \kappa_R.
 \een
The first triplet merely defines, say, constants $C_L^{(a,b,c)}$
as functions of $C_R^{(a,b,c)}$ and $\kappa_{L,R}$. The ratio of
both of these equations eliminates all these constants and leads
to the single complex equation
 \be
   \kappa_L\,{\rm cotan}\, \kappa_L=
 -  \kappa_R\,{\rm cotan}\, \kappa_R.
 \label{rowsyx}
 \ee
Under this constraint our final quantization condition will result
from the insertion of the ansatz (\ref{ansatzf}) in the
Schr\"{o}dinger eq.~(\ref{SE}) at $x \in (-1,0)$,
 \be
 \left (
 \begin{array}{ccc}
  \kappa^2_L+{\rm i}\,Z&{\rm i}\,Y&{\rm i}\,Y\\
 {\rm i}\,Y& \kappa^2_L+{\rm i}\,Z&{\rm i}\,Y\\
 {\rm i}\,Y&{\rm i}\,Y& \kappa^2_L+{\rm i}\,Z
 \ea
 \right )\,
 \left (
 \ba
 C_L^{(a)}\\
 C_L^{(b)}\\
 C_L^{(c)}
 \ea
 \right )
 =
 E\,\left (
 \ba
 C_L^{(a)}\\
 C_L^{(b)}\\
 C_L^{(c)}
 \ea
 \right )
 \label{SEma}
 \ee
and at $x \in (0,1)$,
 \be
 \left (
 \begin{array}{ccc}
  \kappa^2_R-{\rm i}\,Z&-{\rm i}\,Y&-{\rm i}\,Y\\
 -{\rm i}\,Y& \kappa^2_R-{\rm i}\,Z&-{\rm i}\,Y\\
 -{\rm i}\,Y&-{\rm i}\,Y& \kappa^2_R-{\rm i}\,Z
 \ea
 \right )\,
 \left (
 \ba
 C_R^{(a)}\\
 C_R^{(b)}\\
 C_R^{(c)}
 \ea
 \right )
 =
 E\,\left (
 \ba
 C_R^{(a)}\\
 C_R^{(b)}\\
 C_R^{(c)}
 \ea
 \right ).
 \label{SEna}
 \ee
As long as the energies are assumed real, this indicates that we
may put $\kappa_R = s + {\rm i}t = \kappa_L^*$ with, say, positive
$s > 0$ and any real $t \in (-\infty,\infty)$. In this notation
the complex eq.~(\ref{rowsyx}) degenerates to the real implicit
formula
 \be
 2s\,\sin 2s + 2t\,\sinh 2t = 0
 \label{seculara}
 \ee
which first occurred in ref.~\cite{sqw} and which has thoroughly
been studied in ref.~\cite{sgezou}. In our present model,
eq.~(\ref{seculara}) has to be combined with the complex secular
equation
 \be
 \det\,
 \left (
 \begin{array}{ccc}
  (s + {\rm i}t)^2-{\rm i}\,Z-E&-{\rm i}\,Y&-{\rm i}\,Y\\
 -{\rm i}\,Y& (s + {\rm i}t)^2-{\rm i}\,Z-E&-{\rm i}\,Y\\
 -{\rm i}\,Y&-{\rm i}\,Y& (s + {\rm i}t)^2-{\rm i}\,Z-E
 \ea
 \right ) =0.
 \label{secularbe}
 \ee
The resulting three real conditions have to specify the two
unknown real parameters $s=s_n$, $t=t_n$ and the energy $E=E_n$,
$n = 0, 1, \ldots$. Once we set $E = s^2-t^2-\alpha$ (with a real
$\alpha$) and $Z=2st+\beta$ (with a real $\beta$) we may re-write
eq.~(\ref{secularbe}) as a pair of the real polynomial equations
 \be
 \bal
 \alpha^3-3\,\alpha\,
 \left (\beta^2-Y^2
 \right )=0,\\
 \beta^3-3\,\beta\,
 \left (
 \alpha^2+Y^2
 \right )+2\,Y^2=0.
 \ea
 \label{secularcet}
 \ee
In the preliminary test we shall assume that
$\alpha=\alpha_{(tentative)} \neq 0$. From the first row we get
$\alpha^2_{(tentative)}=3\left ( \beta^2-Y^2 \right )$ which
simplifies the second row to the solvable cubic equation with the
three roots,
 \ben
 \beta_1^{(tentative)}=Y, \ \ \ \
 \beta_2^{(tentative)}=\beta_3^{(tentative)}= -\frac{1}{2}\,Y
 , \ \ \ \ \alpha^{(tentative)} \neq 0.
 \een
Their insertion in the definition of $\alpha$ gives the respective
solutions
 \ben
 \alpha_1=0, \ \ \ \
 \alpha_{2,3} = \pm\frac{3 {\rm i}}{2}\,Y
 \een
all of which contradict our assumptions. We are forced to conclude
that we always have the vanishing $\alpha=0$,
 \be
 E = s^2-t^2.
 \label{ansatzfg}
 \ee
At $\alpha=0$ the secular equation (\ref{secularcet}) leads to the
unique triplet of roots
 \be
 \beta_1=-2\,Y, \ \ \ \
 \beta_2=\beta_3= Y
 , \ \ \ \ \alpha = 0.
 \label{rootia}
 \ee
Their respective insertion in eqs.~(\ref{SEma}) and/or
(\ref{SEna}) gives the unnormalized eigenvector
 \be
 \left (C_{(1)}^{(a)},C_{(1)}^{(b)},C_{(1)}^{(c)}\right )
  \sim \left (1^{},1,1
 \right )
 \label{veca}
 \ee
plus the two other, due to the degeneracy, non-unique vectors
available, say, in an orthogonalized representation
 \be
 \left (C_{(2)}^{(a)},C_{(2)}^{(b)},C_{(2)}^{(c)}\right )
  \sim \left (1,-1^{},0
 \right ), \ \ \ \ \ \ \ \
 \left (C_{(3)}^{(a)},C_{(3)}^{(b)},C_{(3)}^{(c)}\right )
  \sim \left (1,1,-2^{}
 \right )
 \label{vecbe}
 \ee
which, incidentally, coincides with the Jacobi-coordinate recipe
for the three equal-mass particles~\cite{low}.

Our knowledge of the roots (\ref{rootia}) enables us to eliminate
one of the real parameters (say, $t$) as lying on one of the two
different hyperbolic curves,
  \be
  t=t^{(\sigma)} (s) = \frac{1}{2s}\,Z_{ef\!f}(\sigma), \ \ \
  \ \ \
  Z_{ef\!f}(1) = Z+2\,{Y}, \ \ \ \ \
  Z_{ef\!f}(2,3) = Z-{Y}\,.
 \label{rooty}
 \ee
Our construction of bound states is completed. In terms of the
parameters $s$ and $t$ they are determined by formulae
(\ref{ansatzf}), (\ref{ansatzfg}), (\ref{veca}) and (\ref{vecbe}).
The parameters themselves must be fixed by the pair of eqs.
(\ref{rooty}) and (\ref{seculara}). In a way described more
thoroughly in ref.~\cite{sqw}, all the physical roots of our
secular eq.~(\ref{seculara}) may be re-parametrized by the formula
 \be
 s=s_n=\frac{(n+1)\pi}{2}+(-1)^n\varepsilon_n
 , \ \ \ \ \ n = 0, 1, \ldots\
 \label{nema}
 \ee
where the new unknown quantities $\varepsilon_n$ remain small not
only in the vicinity of the well known Hermitian case where both
the coupling constants $Y$ and $Z$ remain sufficiently small but
also at all the sufficiently large $n\geq n_0$. Thus, one may
calculate them perturbatively in both these regimes~\cite{Batal}.

\section{The determination of the domain where the energies remain
real}

In the $s-t$ plane we may visualize all the roots $(s_n,t_n)$ as
intersections of the two hyperbolic curves (\ref{rooty}) with all
the family of the $(t \to -t)-$symmetric ovals representing the
complete graphical solution of our second secular implicit
constraint (\ref{seculara}) (cf. \cite{sqw}). As long as  $Y>0$,
all the present energies $E_n$ remain real if and only if
 \be
 Y-Z_{crit}\leq
 Z
 \leq
 Z_{crit}-2Y.
  \label{critin}
 \ee
We may recollect the commentary in ref.~\cite{sgezou} and conclude
that  $Z_{crit} \approx 4.475$ at the present choice of the units
$\hbar = 2m = 1$. In the $(Y,Z)$ plane the condition
(\ref{critin}) is satisfied inside a fairly large triangle with
(approximate) vertices $(0,4.4753)$, $(0,-4.4753)$ and
$(2.9835,-1.4918)$ (cf. Figure 1).

The critical parameter $Z_{crit}$ remains the same for several
different square-well systems. It determines the boundary of the
physical domain in the single-channel square well as well as in
all its classical \cite{Batal} and supersymmetric \cite{sqws}
partners. Still, only its four-digit estimate has been published
up to now~\cite{sgezou}. Moreover, even that improvement of the
original three-digit estimate of ref. \cite{sqw} by one digit did
not seem easy. This apparently discouraged, undeservedly, any
other attempts. For example, an interesting alternative approach
of ref.~\cite{sqwt} using a discretization of the coordinates has
only been studied in the lowest possible approximation giving just
a schematic initial estimate $4\sqrt{2}\approx 5.66$ of
$Z_{crit}$. In fact, a more complicated problem {\em with complex
energies} has been solved during the most successful numerical
attempt in ref.~\cite{sgezou}. For this reason, let us show now
that a systematic improvement of the precision of $Z_{crit}$ can
be made feasible at a reasonable computational cost.

Firstly let us summarize the situation where one locates,
graphically \cite{sqw}, the first two single-channel non-Hermitian
square-well energies $E_{0,1}=s^2_{0,1}-t^2_{0,1}$ as related to
the neighboring intersections $(s_{0,1},t_{0,1})$ of the
implicitly defined $Z-$independent curve $s=s^{(a)}(t)$ [with $
2s^{(a)}(t)\,\sin[ 2s^{(a)}(t)] =-2t\,\sinh 2t $ from
eq.~(\ref{seculara}) above] with the $Z-$dependent but much more
elementary hyperbolic curve $s^{(b)}(t)=Z/(2t)$ of
eq.~(\ref{rooty}). In such a graphical setting it was clarified
that the maximal $Z=Z_{crit}$ at which both the energies $E_{0,1}$
remain real is the point at which they [as well as the neighboring
intersections $(s_{0,1},t_{0,1})$] merge. Thus, the value of
$Z=Z_{crit}$ is defined as a parameter of confluence at which
$E_{0}$ precisely coincides with $E_{1}$.

At this point the curves $s^{(a)}(t)=\pi-\varepsilon(t)$ and
$s^{(b)}(t)$ will osculate at a certain ``intersection" point
$t_{crit}$ and ``exceptional" energy $E_{crit}$. We have to
guarantee that both our curves and both their tangents coincide,
 \be
 \varepsilon(t_{crit}) =\pi-\frac{Z_{crit}}{2t_{crit}},
 \ \ \ \ \ \ \ \ \ \
 \partial_t \varepsilon(t_{crit})
 =
 \frac{Z_{crit}}{2t_{crit}^2}.
 \label{tosolve}
 \ee
The derivative is defined in terms of the positive shift function
$\varepsilon(t)<\pi/4$,
 \ben
 \partial_t \varepsilon^{}(t)=
 \frac{\sinh\,2t +2t\,\cosh \,2t}
 {2\,\left [
 \pi-\varepsilon^{}(t)
 \right ]\,\cos 2\varepsilon^{}(t)-\sin 2\varepsilon^{}(t)
 }
 \een
but the definition of the function $\varepsilon(t)$ itself is
merely implicit,
 \ben
 \sin\left [ 2\,\varepsilon(t)
 \right ]=
 \frac{t\,\sinh\,2t}{\pi-\varepsilon^{}(t)}.
 \een
Fortunately, it may be re-interpreted as a quickly convergent
iterative recipe of a `generalized continued-fraction' type,
 \be
 \varepsilon^{}_{(new)}(t)
 =\frac{1}{2}\,
 \arcsin \left [ 2\,
 \frac{t\,\sinh\,2t}{\pi-\varepsilon^{}_{(old)}(t)} \right ].
 \label{formerec}
 \ee
With the initial $\varepsilon^{}_{(lower)}(t)=\pi/4$ and
$\varepsilon^{}_{(upper)}(t)=0$, the $N-$th iteration of
eq.~(\ref{formerec}) represents the desired {\em explicit}
definition  of the respective {\em approximate} functions
$\varepsilon^{}_{(lower)}(t)$ and $\varepsilon^{}_{(upper)}(t)$.
Their knowledge enables us to solve the two coupled algebraic
equations (\ref{tosolve}) for the two unknown quantities
$t_{crit}$ and $Z_{crit}$ at each particular choice of $N$. The
both-sided convergence of this recipe is illustrated in
Table~\ref{table1}. For the sake of completeness, its last-row
items may be complemented by the corresponding values of
$t_{crit}^{(lower)}=0.839393459 $,
$t_{crit}^{(upper)}=0.839393461$, $s_{crit}^{(lower)}=
2.665799044$, $s_{crit}^{(upper)}= 2.665799069$ and
$E_{crit}^{(lower)}= 6.401903165$ and $E_{crit}^{(upper)}=
6.401903294$.

\section{A remark on the interpretation of the model}

In summary, we felt inspired by several physical applications of
scenario {\bf [a]} which have recently been offered within
relativistic quantum field theory \cite{BCJ} and first-quantized
relativistic quantum mechanics \cite{ANN} as well as in quantum
cosmology \cite{MCJ} and in the classical magnetohydrodynamics
\cite{Uwe}. In these cases one often employs the partitioned and
manifestly Hermitian and involutive $ {\bf P}$, in the latter
three contexts at least \cite{KGja43}. In our present letter we
complemented these studies by an illustration of several new
possibilities emerging within the scenario~{\bf [b]}.

In our present coupled-channel model the Hilbert space is
partitioned in subspaces. In fact, there is no real novelty in
such a procedure of the model-building as the various ${\bf
P}-$pseudo-Hermititian coupled-channel operators are known to
result from the relativistic Sakata-Taketani equations
\cite{FV,FVa} and from their various higher-spin generalizations
and/or non-relativistic analogues~\cite{FN}. A fresh example of a
coupling of channels in non-Hermitian context may be found in our
recent Klein-Gordon study of the influence of certain external
solvable delta-function interactions \cite{KGja87}. The Hermitian
part of our forces was postulated there in the same deep
square-well form of eq.~(\ref{tri}) and only the pseudo-metric
${\bf P}={\bf P}^\dagger$ has been chosen Hermitian, of type {\bf
[a]}.

Let us emphasize that from the point of view of Quantum Mechanics
it is important to know that there exists (at least one)
specification of the scalar product which leads to the positive
definite physical norm. Unfortunately, its {\em explicit
construction} is usually fairly complicated in practice. In this
sense, the exact solvability of the square-well-like models
simplifies significantly the {\em perturbative} construction of
the related ``physical" metric $\Theta \neq {\bf P}$ in Hilbert
space \cite{Batal}. In the other words, our knowledge of $\Theta$
enables us to assert that all the observables in our models
acquire the so called quasi-Hermiticity property \cite{Geyer} and
that only in terms of the (by definition, positive-definite and
self-adjoint) $\Theta$ they may be assigned the standard
probabilistic interpretation \cite{Geyer} -- \cite{Geyere}.

In our present non-relativistic coupled-channel example the
transition to the scenario {\bf [b]} did not influence the methods
of the construction of the physical metric so that they need not
be discussed separately. Even all of their technical details
remain the same for our specific choice of the operator ${\bf P}$
since ${\bf P}^3={\cal P}$. This is an accidental aspect of our
assumptions (\ref{provazanekanaly3}) and (\ref{aliholker}) which
makes the validity of the standard rule (\ref{aliholk}) an
immediate consequence of our assumption (\ref{aliholker}).

In an opposite direction one may say that our present assumptions
concerning the symmetry of the Hamiltonian are {\em stronger}
since our $H$ commutes with the product ${\cal S}=\left [{\bf
P}^{-1}\right ]^\dagger{\bf P} \neq I$. In effect we preserve and
{\em complement} the ``old" eq.~(\ref{aliholk}) by {\em another}
assumption. As long as we {impose more symmetry}, the degeneracy
of some levels occurs. From the practical point of view such a
phenomenon is the consequence of our choice of the complicated
${\bf P} \neq {\bf P}^\dagger$. Of course, a wealth of new
features of the spectrum may be expected to emerge from the more
systematic study of some less schematic non-self-adjoint
``generalized parities" ${\bf P}$ in the future.

\section*{Acknowledgement}

Work supported by the grant Nr. A 1048302 of GA AS CR.


\newpage

\section*{Table captions}

\subsection*{Table 1. Numerical determination of the critical
coupling}

\section*{Figure captions}

\subsection*{Figure 1. Triangular domain of the allowed couplings $Y$ and
$Z$}


   \begin{table}
 \caption{Numerical determination of the critical coupling}
\label{table1}
  \begin{center}
  \begin{tabular}{||c|l|l||}
  \hline\hline
  &&\\
  iteration
 &$Z_{crit}^{(lower)}$&
 $Z_{crit}^{(upper)}$\\
 $N$&&\\
  \hline
\hline
 0&4.299&4.663\\
 2&4.4614&4.4857\\
 4&4.47431&4.47601\\
 6&4.475239&4.475357\\
 8&4.47530381&4.4753119\\
 10&4.475308262&4.475308823\\
 12&4.475308560&4.475308614\\
  \hline
\hline
\end{tabular}
\end{center}
\end{table}


\newpage

\begin{thebibliography}{00}

\bibitem{BBjmp}
C. M. Bender, S. Boettcher and P. N. Meisinger, J. Math. Phys. 40
(1999) 2201.

\bibitem{BB}
C. M. Bender and S. Boettcher, Phys. Rev. Lett. 80 (1998) 4243

\bibitem{ali}
A. Mostafazadeh, J. Math. Phys. 43 (2002)  205.

\bibitem{Bijan}
M. Znojil,
J. Math. Phys. 45 (2004) 4418.

\bibitem{Bijana}
M. Znojil,
J. Math. Phys. 46 (2005) 062109.

\bibitem{Bijanb}
H. B\'{\i}la, V. Jakubsk\'{y}, M. Znojil, B. Bagchi, S. Mallik and
C. Quesne, Czech. J. Phys. 55 (2005) 1075.

\bibitem{Langer}
H. Langer and C. Tretter, Czech. J. Phys. 54 (2004) 1113.

\bibitem{sqw}
M. Znojil, Phys. Lett. A. 285 (2001) 7.

\bibitem{sgezou}
M. Znojil and G. L\'{e}vai,
Mod. Phys. Letters A 16 (2001) 2273.

\bibitem{low}
M. Znojil, J. Phys. A: Math. Gen. 36 (2003) 9929.

\bibitem{Batal}
A. Mostafazadeh and A. Batal,
J. Phys. A: Math. Gen. 37 (2004) 11645.

\bibitem{sqws}
B. Bagchi, S. Mallik and C. Quesne, Mod. Phys. Lett. A 17 (2002)
1651.

\bibitem{sqwt}
S. Weigert, Czech. J. Phys. 55 (2005) 1183.

\bibitem{BCJ}
C. M. Bender, Czech. J. Phys. 54 (2004) 13.

\bibitem{ANN}
A. Mostafazadeh,
Class. Quantum Grav. 20 (2003) 155.

\bibitem{MCJ}
A. Mostafazadeh, Czech. J. Phys. 54 (2004) 93.

\bibitem{Uwe}
U. G\"{u}nther and F. Stefani, Czech. J. Phys. 55 (2005) 1099.

\bibitem{KGja43}
M. Znojil, H. B\'{\i}la and V. Jakubsk\'{y}, Czech. J. Phys. 54
(2004) 1143.

\bibitem{FV}
S. Sakata and M. Taketani, Proc. Phys. Math. Soc. Japan 22 (1940)
757.

\bibitem{FVa}
H. Feshbach and F. Villars, Rev. Mod. Phys. 30 (1958) 24.

\bibitem{FN}
W. I. Fushchych and A. G. Nikitin, Symmetries of equations of
quantum mechanics, Allerton Press, New York, 1994.

\bibitem{KGja87}
M. Znojil, Czech. J. Phys. 55 (2005) 1187.

\bibitem{Geyer}
F. G. Scholtz, H. B. Geyer and F. J. W. Hahne, Ann. Phys. (NY) 213
(1992) 74.


\bibitem{Geyera}
M. Znojil,
 math-ph/0104012
 and
Rendiconti del Circ. Mat. di Palermo, Ser. II, Suppl. 72 (2004)
211.

\bibitem{Geyerb}
R. Kretschmer and L. Szymanowski, quant-ph/0105054 and Phys. Lett.
A 325 (2004) 112.

\bibitem{Geyerc}
C. M. Bender, D. C. Brody and H. F. Jones, Phys. Rev. Lett.  89
(2002) 0270401.

\bibitem{Geyerd}
A. Mostafazadeh,
%
%
arXiv: quant-ph/0310164.

\bibitem{Geyere}
C. M. Bender, D. C. Brody and H. F. Jones, Phys. Rev. Lett. 92
(2004) 0119902 (erratum).

\end{thebibliography}
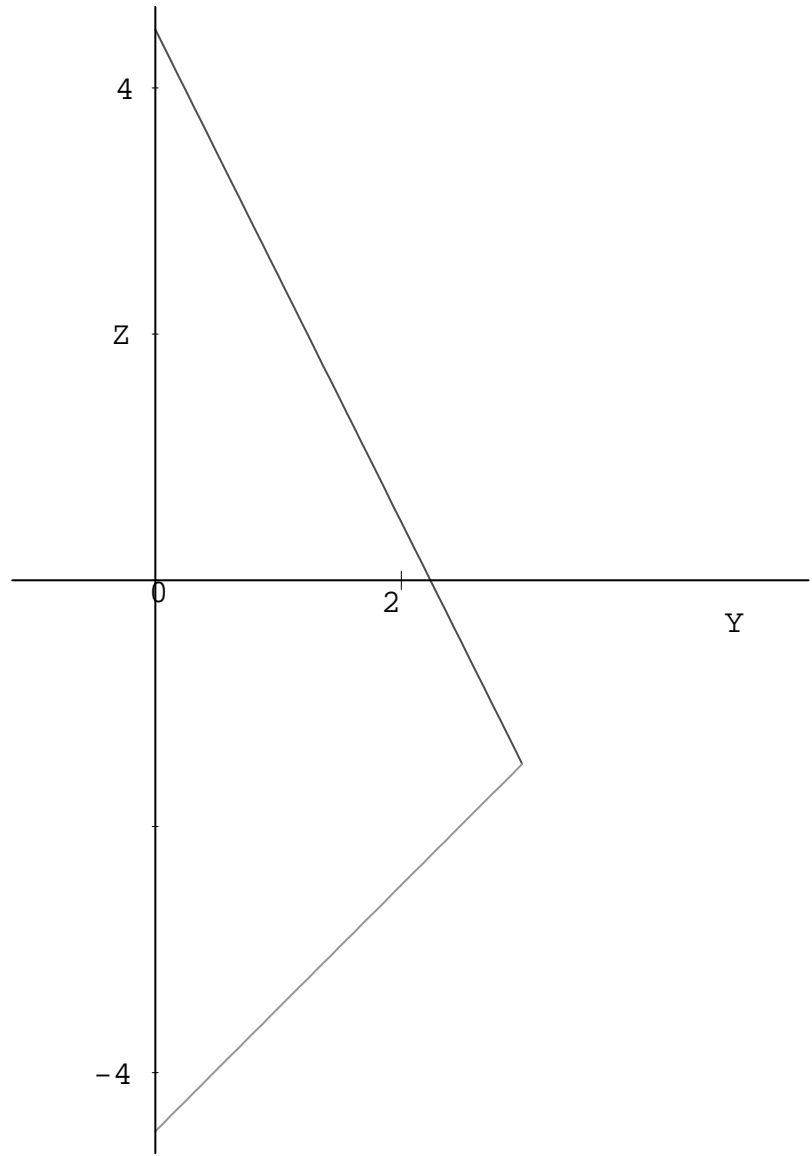
\end{document}